\documentclass[aip, superscriptaddress, preprint]{revtex4}
\usepackage{graphicx}
\usepackage{hyperref}
\hypersetup{
            colorlinks=true,
            linkcolor=blue,
            anchorcolor=blue,
            citecolor=blue
            }
\usepackage{changes}

\usepackage{changes}
\definechangesauthor[name={abc},color=orange]{s.}
\begin{document}

\title
{
Magnetic vortex polarity reversal induced gyrotropic motion spectrum splitting in a ferromagnetic disk
}

\author{Xiaomin Cui}
\email[]{cui.xiaomin.978@m.kyushu-u.ac.jp}
\affiliation{Department of Physics, Kyushu University, 744 Motooka, Fukuoka, 819-0395, Japan}

\author{Shaojie Hu}
\affiliation{Department of Physics, Kyushu University, 744 Motooka, Fukuoka, 819-0395, Japan}

\author{Yohei Hidaka}
\affiliation{Department of Physics, Kyushu University, 744 Motooka, Fukuoka, 819-0395, Japan}

\author{Satoshi Yakata}
\affiliation{Department of Information Electronics, Fukuoka Institute of Technology, 3-30-1 Wajiro-higashi, Higashi-ku, Fukuoka, 811-0295 Japan}

\author{Takashi Kimura}
\email[]{t-kimu@phys.kyushu-u.ac.jp}
\affiliation{Department of Physics, Kyushu University, 744 Motooka, Fukuoka, 819-0395, Japan}


\date{\today}
\begin{abstract}
We investigate the gyrotropic motion of the magnetic vortex core in a chain of a few micron-sized Permalloy disks by electrical resistance measurement with amplitude-modulated magnetic field. We observe a distinctive splitting of the resistance peak due to the resonant vortex-core motion under heightened radio frequency (RF) magnetic field excitation. Our micromagnetic simulation identifies the splitting of the resonant peak as an outcome of vortex polarity reversal under substantial RF amplitudes. This study enhances our understanding of nonlinear magnetic vortex dynamics amidst large RF amplitudes and proposes a potential pathway for spintronic neural computing thanks to their unique and controllable magnetization dynamics.

\end{abstract}


\date{\today}

\maketitle

Magnetic vortices are unique topological magnetic structures found in micron or submicron ferromagnetic elements, characterized by circling in-plane magnetization with a core of perpendicular magnetization\cite{Cowburn1999, Shinjo2000,Choe2004}.
Dynamics based on magnetic vortex include two types, one is the gyrotropic precession of the core in the sub-gigahertz range, and the other one is the core polarization switching to gigahertz steady-state oscillation\cite{Guslienko2002, park2005interactions, Kasai2006, van2006magnetic}.
The fascinating reconfigurable and controllable dynamic properties assure the magnetic vortex of great potential in numerous spin-related applications such as magnetic sensors, spin wave emission sources, and spintronic synapses and neurons for processing, transmitting, and receiving radiofrequency signals\cite{Pribiag2007, kammerer2011magnetic, martins2021non, zhou2021prospect, ross2023multilayer}.
Correspondingly, a study on vortex dynamics earns of great attention. 
The vortex structure gyrates about its equilibrium position with a characteristic eigenfrequency\cite{Guslienko2002, Chen2012}.
By adjusting the shape, the geometrical ratio or the interval distance of the ferromagnetic element containing vortex structure, abundant dynamic properties of the gyration mode could be tailored\cite{park2005interactions, Shibata2004, Barman2010, Vogel2012, Han2013, Yakata2013, Cui2016, lendinez2020temperature}.
In addition to other factors, the resonance's gyration sense is intricately linked to the vortex core polarization. To effectively control this resonance, it's imperative to skillfully manipulate the vortex core polarization\cite{park2005interactions}.
So far, approaches including the utilization of a small AC magnetic field, resonant microwave pulses, DC spin-polarized current or excitation of spin waves have been introduced to reverse the vortex core polarization\cite{van2006magnetic, Pribiag2007, weigand2009vortex, Pigeau2011, kammerer2011magnetic, sushruth2016electrical, warnicke2017tunable}. 
Studies show the origin of vortex core reversal is a gyrotropic field, which is proportional to the velocity of the moving vortex\cite{guslienko2008dynamic}.
 If the core motion speed hits a threshold, the polarity inverts\cite{lee2008universal, sushruth2016electrical}. Several approaches have been demonstrated to reduce the critical velocity for core switching, such as by introducing the nanoscale defects, amplifying the perpendicular anisotropy, or leveraging magnetic interactions between units\cite{fior2016indirect, devolder2019chaos, luo2019edge, mehrnia2021observation}.

Vortex core reversal coincides with deviations from the linear dynamic properties. Nonlinear vortex dynamics often occur at high excitation amplitudes. Investigating novel nonlinear vortex core dynamics not only deepens our understanding of vortex physics but also advances the development of vortex-based spintronic devices. 
This is because the electrical manipulation of nonlinear phenomena provides a significant contribution for advanced functional devices.
Significant theoretical, simulation, and experimental efforts have been made to study the dynamical response of the magnetic vortex in the nonlinear regime, revealing phenomena like fold-over bifurcation, multiphoton resonance, anti-resonances, and translational resonance splitting\cite{Buchanan2007, Gui2009, Guslienko2010}.
As for the mechanism of the nonlinear vortex dynamics, some reports claim that the higher-order terms in magnetostatic potential play a greater role during the large amplitude of vortex gyration mode\cite{Guslienko2010, sukhostavets2013probing, Ding2014}. 
In addition, a large distortion of the vortex core can be effectively excited by high power AC magnetic field with the resonant frequency, leading to nonlinear behaviors \cite{Yamada2007, Vansteenkiste2009}. Although the experimental detection of nonlinear vortex dynamics is crucial for discerning the relationship between vortex core polarity and its electrical response, the constraints of injected RF power in the conventional method based on a monolithic circuit avoid the precise detection of the core dynamics. We have developed a sensitive electrical detection method of the vortex core dynamics by using electrically separated excitation and detection circuits\cite{Cui2015}.
In this research, we extend our sensitive detection technique to investigate the nonlinear vortex dynamics in the chain of the multiple Py disks induced by a high-amplitude AC magnetic field. 
Our observations highlight a resonant peak splitting at elevated RF power when analyzing the power-dependent dynamic response of the magnetic vortex core in gyration mode. Micromagnetic simulations were performed to elucidate this splitting behavior, revealing the role of vortex core polarity switching in inducing resonant peak bifurcation. It proposes one simple way for probing the polarity switching of magnetic vortex, which opens up attractive means for the study of nonlinear magnetization dynamics, and applications such as tunable oscillator\cite{Pribiag2007, de2009bistability, dussaux2010large} and spintronic neural computing\cite{martins2021non,ross2023multilayer}.

Utilizing electron beam lithography and lift-off techniques, We fabricated a chain of micron-sized disks containing 5 Py disks with a thickness of 40 nm. The Scanning Electron Microscopy (SEM) image illustrating the completed device is presented in Fig. 1(a).  
The diameter and edge-edge interval distance of the adjacent disks are 3 $\rm\mu m$ and 2 $\rm\mu m$, respectively. The Cu pad with a thickness of 200 nm and a width of 500 nm was prepared to connect the adjacent Py disks. Here, the Cu pad was up-shifted with a distance of 600 nm from the core of the Py disk. Periodical Cu electrodes with the same thickness as the Cu pad were also prepared on the top of each Py disk for applying AC current. The Py disk and Cu electrodes are electrically isolated through a patterned $\rm SiO_2$ film, precisely tailored to a thickness of 100 nm, ensuring insulation of their electrical connection.

The dynamic properties of the magnetic vortices were detected using lock-in electric measurement technique while the vortex core is excited by the amplitude-modulated Oersted field generated accurately by each Cu electrode\cite{Cui2015,hu2022significant}. 
As shown in Fig. 1(a), during the measurement, an in-plane external magnetic field was applied along the chain of the Py disks to create a nonuniform domain structure by displacing the vortex core from the center.
We flow one DC current of 5 mA in the chained disks and detect the responding voltage of the magnetic vortex device by sweeping the RF frequency. 
Fig. 1(b) shows a representative spectrum measured at an RF power of -20 dBm by sweeping frequency from 50 MHz to 200 MHz under a static magnetic field of 18 mT. A clear resonant peak has been obtained at 126.8 MHz due to the gyrotropic motion of the vortex core.
The resistance change between the resonant peak and the baseline was defined as the effective resistance change $\rm \Delta R_{\rm res}$, which represents the magnetoresistance change between the oscillation and the non-oscillation states of the magnetic vortex core in this device.

\begin{figure}
\begin{center}
\includegraphics[width=6.4in]{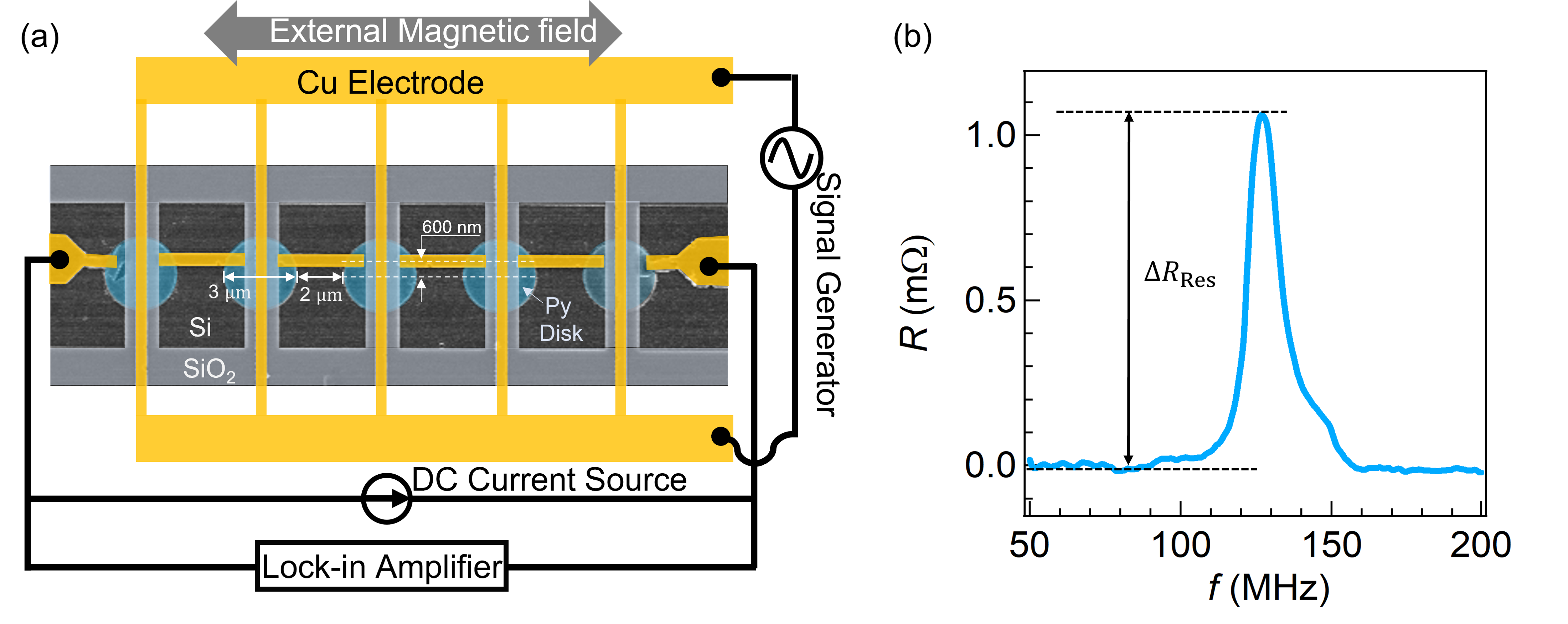}
\end{center}
\caption{(a) Schematic representation of the electrical measurement setup accompanied by an SEM image displaying the fabricated magnetic vortex device; A static magnetic field is applied along the direction of the disk chain. During the measurement, the amplitude of the AC signal was modulated by a low-frequency sinusoidal wave of 1.73 kHz. A DC current was applied to detect the voltage change of the magnetic vortex device. (b) A representative spectrum showing the frequency dependence of the average resistance change of the device with a diameter of 3 $\rm \mu$m and edge-to-edge interval of 2 $\rm \mu$m measured under RF injection power of -20 dBm. Here, the $\rm \Delta R_{\rm res}$ was defined as the difference between the resonant peak and baseline, which represents the magnetoresistance change between the oscillation and the non-oscillation states of the magnetic vortices.}
\end{figure}


In our endeavor to comprehensively analyze the dynamic properties inherent to the chained Py disks, we undertook systematic measurements of the resistance response by sweeping the RF frequency under different external magnetic fields.  Figures 2(a) and 2(b) provide a visual representation in the form of color images, illustrating the device resistance as a function of both microwave frequency and the imposed external magnetic field under RF power of -20 and -10 dBm, respectively. 
A notable observation from Fig. 2(a) is the emergence of a distinct 'M' shaped profile characterizing the resonance frequency when plotted as a function of the external magnetic field at an RF power of -20 dBm. 
The data derived from this figure suggests that the external magnetic field has a relatively subdued modulatory influence on the resonance frequency.
In stark contrast, Fig. 2(b) manifests a divergence from this pattern. At an RF power of -10 dBm, the resonance frequency appears to adopt a dual 'M' configuration. This intriguing shift suggests a pronounced splitting in the resonance peak of the spectrum. Furthermore, the frequency disparity between these two 'M' shaped configurations appears markedly greater compared to the results influenced solely by field modulation. To delve deeper into these nuanced differences and glean more granular insights, we selected a range of magnetic fields. The resultant frequency-dependent spectra derived from these selections are comprehensively illustrated in Fig. 2(c) and 2(d).
In our examination of the device's response, a discernible transformation in the Lorentzian-like spectrum is evident. This transformation progresses from a peak (observed at -20 mT) to a  dip (registered at 4 mT) and ultimately reverts to a peak (noted at 17 mT). This progression is interspersed with transitions through anti-symmetric Lorentzian-like spectra, specifically observed at magnetic field intensities of 1 mT and 14 mT. The intricate dynamics behind this transition behavior can be ascribed to the atypical resistance alterations contingent upon the central positioning of the vortex core. This phenomenon has been extensively detailed and rationalized in our recent work\cite{hu2022significant}.
Furthermore, when we modulate the RF power, elevating it from -20 dBm to a higher -10 dBm, we observe a nuanced alteration in the spectral characteristics. The once singular resonant peak (or dip, as the case may be) within the Lorentzian-like spectrum bifurcates, resulting in the emergence of two distinct resonant peaks (or dips). It's imperative to note, however, that this spectral division predominantly manifests in proximity to the dip of the anti-symmetric Lorentzian-like spectrum, as opposed to its peak.

\begin{figure}
\begin{center}
\includegraphics[width=6.4in]{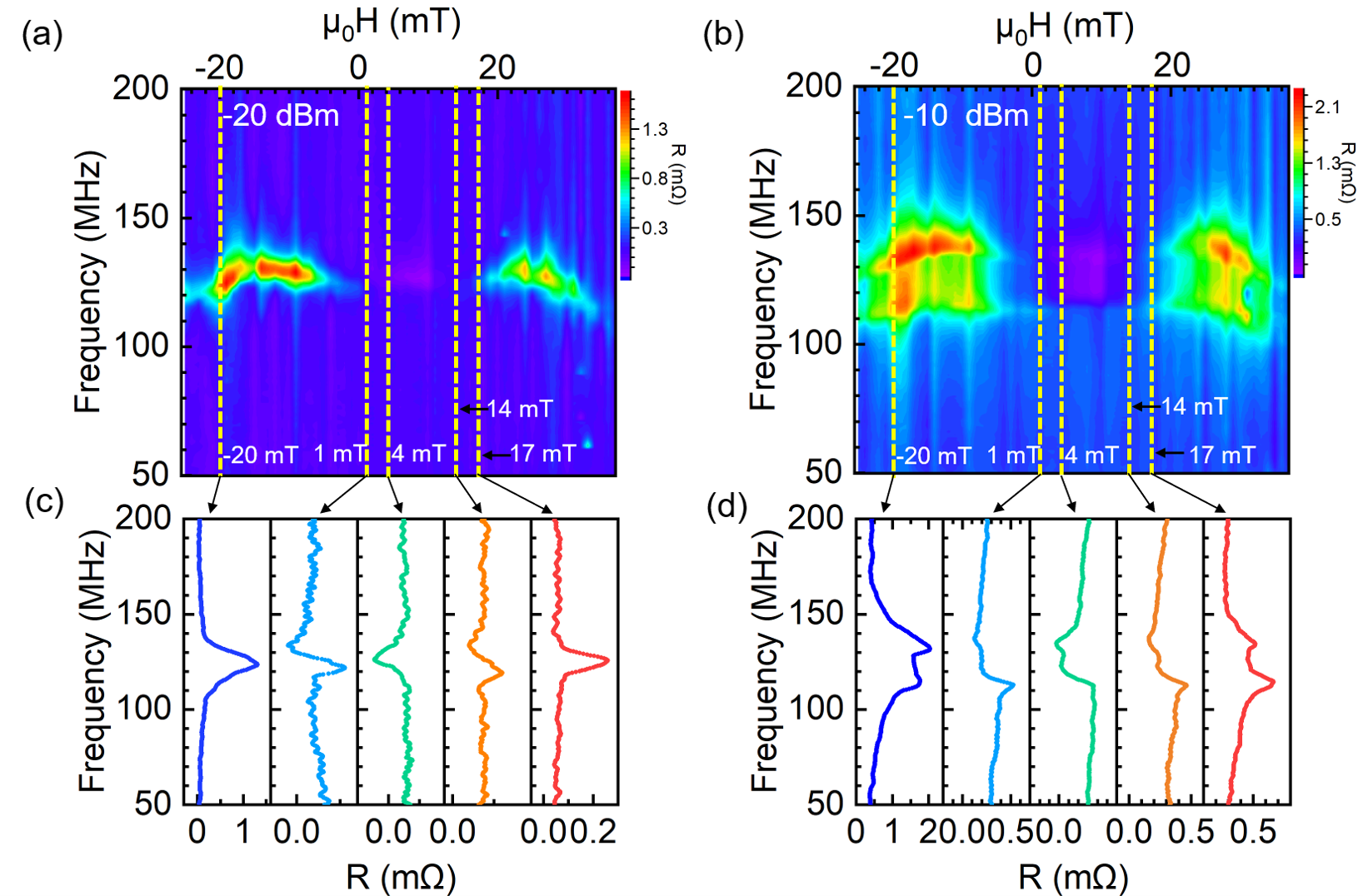}
\end{center}
\caption{
(a) Experimental observed dynamic spectra of the chained Py disks at different external magnetic fields and RF power of -20 dBm.
(b) Experimental observed dynamic spectra of the chained Py disks at different external magnetic fields and RF power of -10 dBm.
(c) The resistance as a function of the input RF frequency under various external fields marked as the dotted lines in (a).
(d) The resistance as a function of the input RF frequency under various external fields marked as the dotted lines in (b).
}
\end{figure}

In an endeavor to delve deeper into the intricacies of the observed splitting behavior, our research team undertook meticulous measurements, examining the power dependence of the magnetic vortex dynamics. This was systematically conducted across an RF amplitude spectrum ranging from -33 dBm to 0 dBm, under the magnetic fields specified earlier in our discourse.
The graphical representation in Fig.3(a), using a color image plot, offers a clear depiction of the resonant frequency's behavior.  Specifically, within magnetic field environments of -20 mT and 17 mT, a singular resonant peak manifests a discernible bifurcation, evolving into two separate resonant peaks as the RF amplitude augments. 
It's pivotal to highlight that this augmentation in RF amplitude invariably leads to an expansion in the frequency separation interposed between these two emergent resonant peaks. Furthermore, as one continues to elevate the RF amplitude, the extent of this aforementioned separation between the two resonant peaks amplifies correspondingly.
Notably, as the power escalates to -5 dBm the differential between the two peaks culminates at an impressive 37.2 MHz at a magnetic field of 17 mT. For perspective, juxtaposing this with the resonant frequency pinpointed at an RF amplitude of -20 dBm reveals a frequency modulation on the order of approximately 29.6\%. This substantial figure underscores the profound influence exerted by RF power on frequency modulation.


To elucidate the power-dependent dynamics intrinsic to the concatenated Py disks, the spectral data acquired at a magnetic field strength of 17 mT were chosen for more intricate scrutiny, owing to their distinctive power-amplified splitting behavior.
The analysis was conducted at an excitation frequency of 125.6 MHz, a resonant frequency empirically determined at a power level of -20 dBm and a magnetic field strength of 17 mT. Magnetic resistance values at the aforementioned excitation frequency (R) and the baseline resistance ($\sl \rm R_0$) were extracted and employed to calculate the difference in resistance $\sl \Delta \rm R$. The collated data is presented in Fig.3(b), delineated as a function of radio-frequency (RF) power. Notably, both $\sl \rm R_0$ and R demonstrate an exponential augmentation as discerned from Figure 3(b), which suggests a potential correlation with the thermal effects engendered by the escalation in RF power. Additionally, $\sl \Delta \rm R$ exhibits an ascending trajectory, culminating at a maximum value at an RF power level of -19 dBm. Subsequently,  $\sl \Delta \rm R$  begins to exhibit a decremental trend concomitant with the continued escalation of radio-frequency (RF) power.  The underlying mechanisms contributing to this decline in $\sl \Delta \rm R$ at elevated RF power levels remain as yet undetermined. To clarify this phenomenon, we propose to employ micromagnetic simulation techniques. Establishing a correlative framework between the empirical findings and the dynamic processes gleaned from micromagnetic simulations could illuminate the origin of both the power-dependent splitting behavior and the attenuation of $\sl \Delta \rm R$.

\begin{figure}
\begin{center}
\includegraphics[width=6.4in]{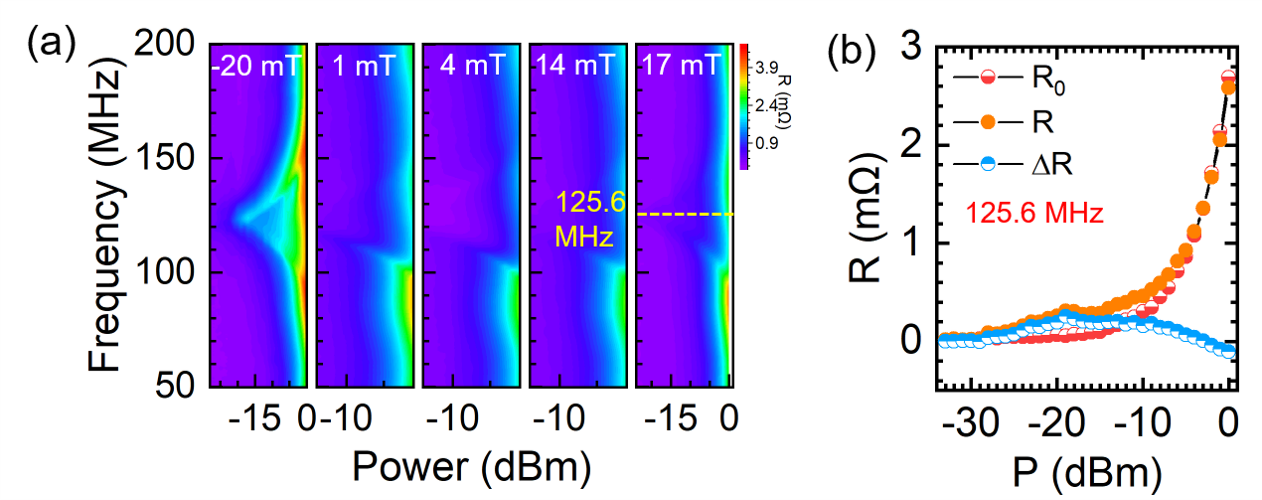}
\end{center}
\caption{ 
(a) Experimentally observed spectra of concatenated Py disks at magnetic fields as a function of RF power across varying magnetic fields of -20 mT, 1 mT, 4 mT, 14 mT and 17 mT.
(b) 
Resistance differential $\sl \Delta \rm R$ between oscillatory and non-oscillatory states as a function of excitation power, evaluated at a magnetic field strength of 17 mT: pertinent to a magnetic vortex excited at 125.6 MHz, the resonant frequency determined at an RF power level of -20 dBm.
}
\end{figure}

We performed the micromagnetic simulation using MuMax$^3$ here. \cite{Vansteenkiste2014}  
The computational domain was architecturally congruent in both dimensions and thickness with the fabricated device. Discretization of this domain was executed utilizing mesh dimensions of 4 nm $\times$ 4 nm $\times$ 40 nm.
We employed canonical microstructural parameters pertinent to Permalloy, featuring an
exchange stiffness constant $\rm A_{ex}= 1.3\times10^{-11}$ $\rm J/m$, damping parameter $\alpha=0.006$, saturation magnetization $\rm M_s=8\times 10^5$ $\rm A/m$, and a null magnetocrystalline anisotropy constant.
To access the observed unique spectra, the dynamic processes were investigated by calculating the corresponding anisotropic magnetoresistance (AMR) in the Py disk. The AMR of Py could be calculated by considering the relation of  $\rho=\rho_0(1+\eta cos^2\theta)$. The resistivity of Py ($\rho_0$) is $3.6\times 10^8\ \Omega \cdot m$. The AMR ratio $\eta$ is about 1.4\%. $\theta$ is the angle between the magnetization and current density direction in the magnetic unit cell. 
Assuming the current density is in the x-direction, the resistivity of each unit cell can be determined based on its magnetization direction using the AMR effect. The overall resistance of the Py disk can calculated by connecting all the resistor units in parallel and series for each magnetic state. 
To corroborate our empirical findings, we intend to execute micromagnetic simulations, selecting radio-frequency (RF) power levels of -20 dBm and -10 dBm for comparative analysis. Prior to this, the resonant frequencies at a magnetic field strength of 17 mT were calculated via micromagnetic simulations. Subsequently, dynamic simulations were conducted at an excitation frequency of 123 MHz. Figure 4(a) delineates the core trajectories of the singular vortex structure at RF power levels of -20 dBm and -10 dBm, superimposed on a contour map depicting magnetoresistance within the Permalloy disk. Time-dependent anisotropic magnetoresistance, calculated under the constraint of uniform current density, is presented in Fig. 4(b) and 4(c)\cite{cui2014detection,hu2022significant}. Remarkably, at an RF power of -20 dBm, the vortex core manifests stable oscillatory behavior. For the core trajectory under -10 dBm, directional movements are indicated by yellow arrows, elucidating the vortex core's oscillatory mechanics. The vortex core commences rotation from an inferior position in a clockwise direction towards point P1, whereupon it transitions into a counterclockwise rotation beginning at point P2, thus illustrating a polarity reversal\cite{guslienko2008dynamic}. To substantiate this observation, magnetization texture snapshots at specific points were captured, as detailed in Figure 4(d). (Refer to the supplementary video for a complete overview of the simulation duration.)
The magnetization profiles at points P1, P2, and P3 validate the polarity transition from a downward to an upward orientation. Subsequent to this, the vortex core reverts to a clockwise rotational direction as it translocates from point P4 to point P5, coinciding with an additional polarity inversion. Intriguingly, such polarity reversals are conspicuously absent at an RF power of -20 dBm. Collectively, these results insinuate a salient correlation between the vortex polarity reversals and variations in $\sl \Delta \rm R$ contingent on the applied RF power.

\begin{figure}
\begin{center}
\includegraphics[width=6.4in]{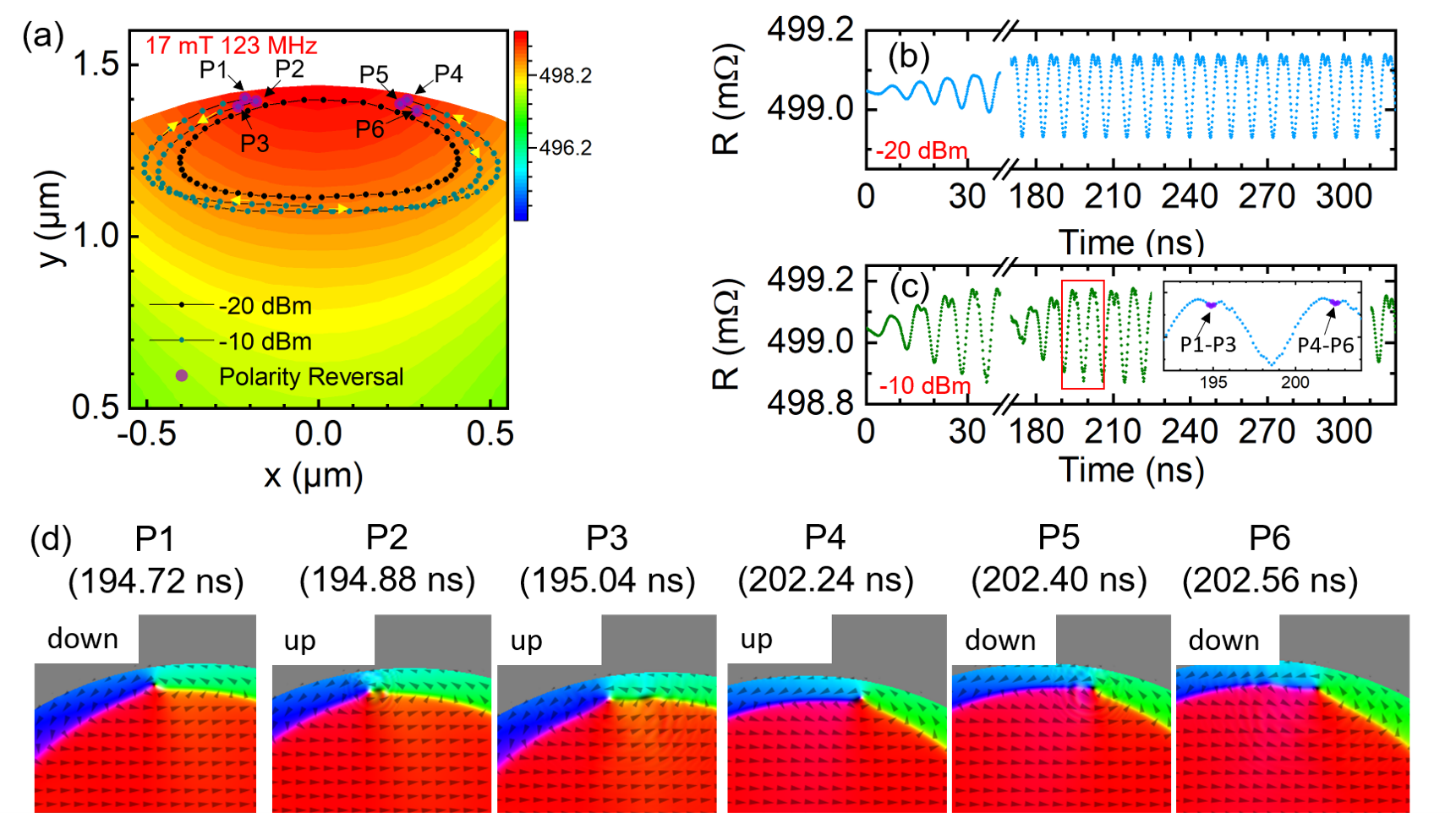}
\end{center}
\caption{ 
(a) The micromagnetic calculated core trajectories of the vortex in the contour map of the Py disk, while it was excited under external field of 17 mT, frequency of 123 MHz and power of -20 dBm and -10 dBm, respectively. Yellow arrows stand for the moving direction of the vortex core during oscillation. The purple dots correspond to the vortex polarity reversal.
(b) and (c) represent the time-dependent resistance of the Py disk excited under -20 dBm and -10 dBm, respectively.
(d) Snapshots of numerical domain structures with the vortex core reversal process corresponding to the highlighted purple points P1 to P6 from (a) and the inset of (c). 
}
\end{figure}

Furthermore, magnetic resistance values were extracted for the non-oscillatory state at t = 0 as well as for the average magnetic resistance during the oscillatory state across 5-10 periods. Subsequently, the difference in resistance between the non-oscillatory and oscillatory states, denoted as $\sl \Delta \rm R$ was calculated across a range of excitation frequencies. The variation of $\sl \Delta \rm R$  as a function of RF frequency is depicted in Figure 5(a). Simulation results reveal a singular resonant peak at 122 MHz under an RF amplitude of -20 dBm. Upon increasing the RF amplitude to -10 dBm, this resonant peak bifurcates into two separate peaks with frequencies of 119.5 MHz and 124 MHz. Further amplification to -5 dBm results in greater frequency splitting, as corroborated in Figure 5(a). These findings are in close accord with the previously mentioned experimental observations. 
As for $\sl \Delta \rm R$ as a function of RF power, depicted in Figure 5(b), the behavior of $\sl \Delta \rm R$ is consistent with the experimental outcomes outlined in Figure 3(b). Notably, the maximum $\sl \Delta \rm R$ value is attained at -16 dBm, slightly exceeding that of the experimental results. This discrepancy may be attributed to defects induced during the device fabrication process, where the critical value is achieved at a relatively lower energy level. 
When the excitation frequency for the magnetic vortex aligns with or is proximal to, the resonant frequency, $\sl \Delta \rm R$ commences a decline post the critical value. Conversely, when the frequency deviates significantly from the resonant frequency, the vortex core experiences minimal oscillation, thereby rendering the RF power enhancement less impactful on $\sl \Delta \rm R$. In essence, the reduction of 
$\sl \Delta \rm R$ at the resonant frequency, occasioned by the reversal of vortex polarity, engenders the observed splitting in the dynamic spectra. Additionally, higher power levels induce a greater incidence of vortex polarity switching, further contributing to the reduction of $\sl \Delta \rm R$.

\begin{figure}
\begin{center}
\includegraphics[width=6.4in]{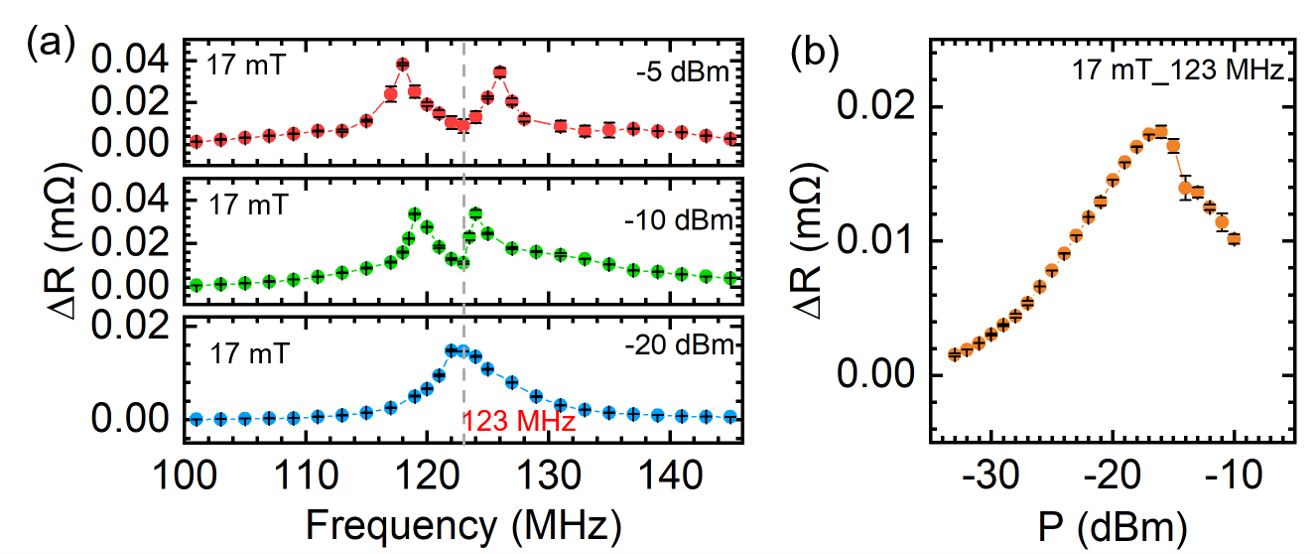}
\end{center}
\caption{ 
(a) Calculated $\sl \Delta \rm R$ as a function of the excitation frequency under the external field of 17 mT and excitation power of -20 dBm, -10 dBm and -5 dBm, respectively.
(b) Calculated $\sl \Delta \rm R$ with respect to the input RF power under the external field of 17 mT and frequency of 123 MHz.
}
\end{figure}

In conclusion, by employing an amplitude-modulated magnetic field to accurately stimulate the gyroscopic motion of the magnetic vortex core in each disk, we have elucidated the splitting phenomena of the resonant peak in magnetic vortices under elevated RF amplitudes. This investigation was facilitated through the exploration of power-dependent resistance spectra. We observed that the frequency separation between the dual resonant peaks magnifies concomitantly with increasing RF amplitude. Utilizing micromagnetic simulations, we successfully replicated the experimental observations and substantiated that this resonant peak splitting arises due to vortex polarity reversal at high RF amplitudes. These empirical findings contribute to an enriched comprehension of nonlinear vortex dynamics under substantial excitation amplitudes and provide a potential avenue for the advancement of tunable microwave-assisted spintronic devices or spintronic neural networks.

\section*{Acknowledgement}
This study is partially supported by JSPS KAKENHI Grant Numbers 17H06227, 21H05021 and JST CREST (JPMJCR18J1).

\section*{Additional information}
Supplementary information shows the video of the magnetization texture of the Py disk observed under 17 mT and -10 dBm and excitation frequency of 123 MHz as a function of the simulation time.

\section*{Data Availability}
The data that support the findings of this study are available from the corresponding author upon reasonable request.

\bibliography{splitting} 
\bibliographystyle{rsc} 

\end{document}